\documentclass[%
 reprint,
 superscriptaddress,
 amsmath,amssymb,
 aps,
 prb
]{revtex4-2}

\usepackage{graphicx}
\usepackage{dcolumn}
\usepackage[colorinlistoftodos]{todonotes}
\usepackage{siunitx}
\usepackage{graphicx}
\usepackage{lipsum}
\usepackage{todonotes}
\usepackage{float} 
\usepackage{commath}

\begin{document}

\title{Fractal structure, depinning, and hysteresis of dislocations in high-entropy alloys}%

\author{Hoa Thi Le}
\affiliation{Department of Microsystems Engineering, University of Freiburg, Georges-K\"ohler-Allee 103, 79110 Freiburg, Germany}

\author{Wolfram G. Nöhring}
\affiliation{Department of Microsystems Engineering, University of Freiburg, Georges-K\"ohler-Allee 103, 79110 Freiburg, Germany}

\author{Lars Pastewka}
\email[Corresponding author: ]{lars.pastewka@imtek.uni-freiburg.de}
\affiliation{Department of Microsystems Engineering, University of Freiburg, Georges-K\"ohler-Allee 103, 79110 Freiburg, Germany}
\affiliation{Cluster of Excellence livMatS, Freiburg Center for Interactive Materials and Bioinspired Technologies, University of Freiburg, 79110 Freiburg, Germany}

\date{\today}

\begin{abstract}
High-entropy alloys (HEAs) are complex alloys containing multiple elements in high concentrations.
Plasticity in HEAs is carried by dislocations, but the random nature of their composition pins dislocations, effectively hindering their motion.
We investigate the resulting complex structure of the dislocation in terms of spatial correlation functions, which allow us to draw conclusions on the fractal geometry of the dislocation.
At high temperature, where thermal fluctuations dominate, dislocations adopt the structure of a random walk with Hurst exponent $1/2$ or fractal dimension $3/2$.
At low temperature we find larger Hurst exponents (lower dimensions), with a crossover to an uncorrelated structure beyond a correlation length.
These changes in structure are accompanied by an emergence of hysteresis (and hence pinning) in the motion of the dislocation at low temperature.
We use a modified Labusch/Edwards-Wilkinson-model to argue that this correlation length must be an intrinsic property of the HEA.
This means dislocations in HEAs are an individual pinning limit, where segments of the dislocation are independently pinned by local distortions of the crystal lattice that are induced by chemical heterogeneity.
\end{abstract}


\maketitle


\section{\label{sec:level1}Introduction}

High-entropy alloys (HEAs) are complex concentrated alloys (CCAs) containing five or more elements in high concentrations~\cite{Yeh2004NanostructuredOutcomes, Cantor2004MicrostructuralAlloys}.
HEAs obtain their name from their high entropy of mixing, which has been argued to stabilize random solid solutions even in low temperature crystalline phases.
They are compositionally disordered, but unlike metallic glasses~\cite{Hufnagel2016DeformationExperiments}, they retain long-ranged crystalline order.
HEAs can exceed the performance of conventional alloys for structural applications.
They show high strength, toughness and hardness, wear resistance, and they can retain mechanical properties even at high temperatures.

Studying the solute strengthening of HEAs has led to a conceptual understanding of these alloys' mechanical behavior.
Due to local chemical disorder (also referred to as quenched disorder), which originates in the random distribution of different elements, the interaction energy between the dislocation and its environment fluctuates, resulting in pinning of the dislocation.
A predictive theory for solute strengthening in face-centred cubic (FCC) HEAs alloys assumes that dislocations become pinned in regions with energetically favourable fluctuations, as they adopt a wavy configuration to reach a local minimum in potential energy~\cite{Varvenne2016TheoryAlloys,Varvenne2017SoluteAlloys,Utt2022TheAlloys}.
At finite temperatures, thermal disorder is superimposed onto chemical disorder, effectively smoothing the (quenched) disorder field.

Much of our knowledge about dislocations in HEAs has been derived from simple conceptual models.
The simplest model for the time-evolution of a dislocation described by its displacement profile $h(x,t)$ is a Labusch~\cite{Labusch1972-ow,Labusch1981-ic} or Edward-Wilkinson (EW) formulation~\cite{Edwards1982TheAggregate}, sometimes also called minimal line-tension model.
In this model, the free energy of the dislocation is given by
\begin{equation}
    A_\text{EW} = \int_0^L \dif x \, \left\{\frac{\Gamma}{2} \left|\frac{\partial h}{\partial x}\right|^2+U_\text{p}(x, h) - f h\right\},
    \label{eq:qew}
\end{equation}
where $\Gamma$ is the line tension, $U_\text{p}(x, h(x))$ is a random line-energy representing the quenched disorder that the dislocation sees, $f$ is a driving force and $L$ the nominal length of the dislocation.
Canonical disorder models employ a correlation length $\xi$ beyond which the random field $U_\text{p}(x, h)$ is uncorrelated.
For real HEAs, this correlation length should be on the order of the atomic spacing, unless constituent atoms themselves form correlated structures~\cite{Zhang2017-zy,Li2019-gi}.

In the weak pinning limit, EW models show~\cite{Zapperi2001DepinningInteractions} a depinning stress that is controlled by an additional length scale, the Larkin length $\lambda_\text{p}$~\cite{Larkin1970-gy,Labusch1970AHardening,Labusch1972-ow,Larkin1979-rv,Labusch1981-ic}, below which the dislocation is essentially rigid.
The dislocation then sees an obstacle field that is averaged over a length $\lambda_\text{p}$. 
Simulations of this model have yielded power-law correlations with a Hurst~\cite{Hurst1951Long-TermReservoirs,Mandelbrot1982-wd} exponent of $H=1$ at small scales and uncorrelated noise at large scales.
The crossover point depends on applied stress and appears to diverge near depinning~\cite{Zhai2019PropertiesDisorder}.
This behavior is identical to what has been observed in Discrete Dislocation Dynamics (DDD) simulations for dislocations in obstacle fields~\cite{Bako2008DislocationSteel}.
Rodney and coworkers have argued that this cross-over length is essentially the Larkin length $\lambda_\text{p}$~\cite{Rodney2024-at}.
We note that adding nonlinearities to Eq.~\eqref{eq:qew} affects the Hurst exponent.
For example, the Kardar-Parisi-Zhang (KPZ) equation~\cite{Kardar1986-ro} appears to yield an exponent of $H\sim 2/3$~\cite{Ferrero2021CreepLandscape}.

Thermal fluctuations can be included in Eq.~\eqref{eq:qew} by adding random (Langevin-like) noise, but this has received much less attention.
Without quenched disorder, the dislocation adopts a configuration whose displacement power-spectrum reflects the Fourier-transform of the Green's function~\cite{Campana2006-mj} of Eq.~\eqref{eq:qew}, which scales $\propto q^{-2}$ with wavevector $q$ and looks like a Hurst exponent of $H=1/2$~\cite{Zhai2019PropertiesDisorder}.
We note that the actual depinning force (and whether depinning occurs) depends on the distribution of the random pinning or obstacle field~\cite{Coville2010-gc,Dondl2017,Dondl2022}, but much of the literature appears to implicitly assume that this follows Gaussian statistics.

The structure of dislocations in HEAs was also studied in atomistic calculation.
Péterffy and collaborators~\cite{Peterffy2020LengthSolutions} studied edge dislocations in Fe$_{0.70}$Ni$_{0.11}$Cr$_{0.19}$ alloys between $5$~K and $200$~K, while more recently Esfandiarpour and colleagues~\cite{Esfandiarpour2022EdgeDepinning} have looked at a wider range of alloys, all using embedded-atom and modified embedded-atom potentials~\cite{Muser2023-zb}. 
Both studies concluded that the movement of dislocation through a random solid solution is well described using the statistical theory of elastic manifolds by random fields~\cite{Fisher1998CollectiveEarthquakes,
Chauve2000CreepMedia, Rosso2003DepinningManifolds, Ferrero2021CreepLandscape}.
Consistent with results from EW models, dislocation lines form a self-affine structure with a Hurst exponent between $1/2$ and $1$ at short distances, transitioning to uncorrelated noise $(H\sim 0)$ beyond a correlation length $\lambda_\text{p}$, which increases with stress and diverges at the depinning threshold.
Following the arguments put forth for the EW equations, this length $\lambda_\text{p}$ looks like the Larkin length.

Here, we revisit the relationship between depinning and the structure of dislocations in high-entropy alloys, with a particular focus on the combined effects of temperature and compositional disorder.
We investigate the roughness profiles of edge dislocations in an equicomposition FeNiCrCoCu high-entropy alloy and compare it to an unary crystal interacting via a mean-field representation of the alloy, also called an ``average alloy'' (AA) model~\cite{Varvenne2016}.
This unary crystal in the AA model has identical elastic and thermodynamic properties~\cite{Nohring2016ThermodynamicAlloys} but lacks the quenched disorder of the full HEA.
This allows us to disentangle the effects of quenched and thermal disorder on the same material system.
Additionally, we examined the roughness profiles of edge dislocations in FeNiCr and NiCrCo equicomposition CCAs, without applying the AA model to these alloys.
We confirm that under the effect of quenched disorder, the profile of the dislocation has a power-law correlation with Hurst exponent $\approx 2/3$ at short distances.
The exponent crosses over to $1/2$, indicative of a random walk, as the temperature increases.
The crossover to the thermal exponent of $1/2$ is accompanied by a disappearance of dislocation pinning, as directly evidenced by a disappearing hysteresis between forward and backwards movement of the dislocation.

\section{Methods} \label{method}

We simulated the fluctuation of edge dislocations in an equicomposition, face-centered cubic (fcc) FeNiCrCoCu HEA, FeNiCr and NiCrCo CCAs using molecular dynamics (MD) simulations.
The simulations used the embedded-atom method (EAM) potential~\cite{Daw1984Embedded-atomMetals} described in Ref.~\cite{Deluigi2021SimulationsResistance} for FeNiCrCoCu, in Ref.~\cite{Zhou2018-FeNiCr} for FeNiCr and in Ref.~\cite{Li2019-NiCrCo} for NiCrCo. These EAM potentials were specifically designed for HEAs, prevent phase-separation of the components and are stable at high temperatures.
We also constructed an average-representation of the FeNiCrCoCu HEA system, which is a crystal with elastic and thermodynamic properties identical to the HEA, but consisting of a single virtual, average chemical element~\cite{Varvenne2016}.

\begin{figure}[t]
\centering
 \includegraphics[clip,width=\columnwidth]{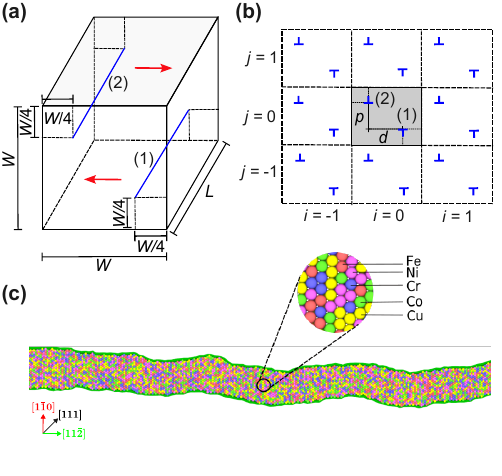}%
\caption{\label{fig:sim_box}\textbf{Illustration of the simulation setup.} \textbf{(a)} Schematic diagram of the simulation cell, showing a dipole consisting of two perfect edge dislocations (1) and (2) in blue. \textbf{(b)} The geometry of two-dimensional periodic dislocation dipole arrays. Two perfect edge dislocations are separated by a horizontal distance $d$ and vertical distance $p$. \textbf{(c)} In our molecular simulation, the edge dislocation splits into two Shockley partials. The snapshot shows the atoms that are not in an fcc environment, color-coded by elements of the FeNiCrCoCu equicomposition HEA. The two partial dislocations are connected by a stacking fault and adopt a wavy configuration.} 

\end{figure}

Our simulation cells are periodic and contain a dipole of edge dislocations (1) and (2), as schematically shown in Fig.~\ref{fig:sim_box}a.
We created three simulation cells with dislocations of length $L=\SI{1500}{\angstrom}$ oriented in the $[1 1 \bar{2} ]$ direction.
The three cells differ by their periodic size in the two perpendicular directions, or equivalently by the distance between the two dislocations in the dipole.
We used cell sizes of $350$, $700$ and $1500$~Å in the $[1 \bar{1} 0 ] $ and $[1 1 1]$ directions.

Simulation cells under periodic boundary conditions are infinitely repeated.
In Fig.~\ref{fig:sim_box}b, we illustrate a two-dimensional periodic dislocation dipole array. The grey box is the original unit cell, which contains two primary edge dislocations (1) and (2) with opposite Burgers vectors.
Under periodic boundary conditions, the unit cell is infinitely repeated as dashed boxes, identifiable by indices $i,j=-\infty,...,+\infty$.
Note that we chose a (periodic) dipole configuration instead of the (more common) single dislocation in a box with rigid boundaries because we can quantify the interaction of the two dislocations within the dipole analytically.
Conversely, the influence of rigid boundaries on the long-ranged displacement field surrounding a single dislocation is more subtle and can be harder to quantify.
We will revisit to the dislocation-dislocation interaction when discussing the results in Sec.~\ref{sec:discussion}.

Our MD simulations followed two protocols: a slow quench protocol, where systems were first equilibrated for $200$~ps at $1025$~K before being cooled to the target temperature at a rate of $10^{11}$~K~$\text{s}^{-1}$, and the immediate quench protocol, where systems were initialized and equilibrated directly at the target temperatures without the intermediate high-temperature step.
We refer to the dislocations obtained in the slow quench as ``well-equilibrated''.
We relaxed the three diagonal components of the stress tensor independently to zero in the final simulation box in $200$~ps runs with a Berendsen barostat~\cite{Berendsen-1984} at the target temperature.
We used a Langevin thermostat during relaxation. 

In the immediate quench protocol, we generated simulation cells with a straight edge dislocation dipole and directly equilibrated for $20$~ps at the target temperatures (1000~K, 400~K, and 5~K) without the preceding slow quench. 
We use this protocol to study the inherent structure of the dislocation~\cite{Stillinger1984PackingSolids}, which is the local potential energy minimum that is closest to a specific atomic configuration.
We used a conjugate-gradient minimization of the potential energy to search for this local minimizer after the short equilibration period.

Well-equilibrated dislocation dipoles were subsequently deformed in simple shear at a strain rate of $10^{7}\,\text{s}^{-1}$.
During shear, we employed a Nosé-Hoover~\cite{Nose1984AEnsemble} thermostat and an Andersen-barostat~\cite{Andersen1980MolecularTemperature} that keeps the three diagonal components of the stress tensor near zero.
Shear stress and the dislocation structure were recorded in strain increments of $0.05\%$.

Our edge dislocations split into two Shockley partials (see Fig.~\ref{fig:sim_box}c).
We used the dislocation extraction algorithm (DXA)~\cite{Stukowski2010ExtractingData} to obtain a representation, $h(x)$, of the displacement of each of the four partial dislocations in our dipole system.
We then used a spatial correlation function to characterize the geometry of the dislocation.
Our primary tool is the displacement-difference autocorrelation function, in its general form given by
\begin{equation}
    R_\alpha(\ell)=\left\langle \left|h(x+\ell) - h(x)\right|^\alpha\right\rangle_x^{1/\alpha},
    \label{eq:acf}
\end{equation}
where the average $\langle \cdot \rangle_x$ is taken over position $x$.
Note that $R_2$ is typically called the ``autocorrelation'' function, and $R_1$ is the structure function.
Different values for $\alpha$ are useful to detect deviations from Gaussian fluctuations, but we here focus on $R_2$ that is the standard deviation of fluctuations.
Another common statistical analysis method that we also use is the power spectral density (PSD).
The dislocation line $h(x)$ is periodic with nominal length $L_y$ and can be expressed as a Fourier series
\begin{equation}
    h(x) = \sum_{n} C_n e^{iq x},
\end{equation}
where $q = 2\pi n/L_y$ is the wavenumber.
The PSD is then obtained as 
\begin{equation}
    C(q_n) = L_y|C_n|^2.
\label{eq:psd}
\end{equation}
Note that $R_2$ and $C$ are related by a Fourier-transform, also known in parts of the physics literature as the Wiener–Khinchin theorem~\cite{Wiener1930-qh,Khintchine1934-sc}.

The final correlation functions are averaged over all four partials and finite-temperature (Nosé-Hoover dynamics) simulations of $10$~ps duration at a fixed simulation cell.
Specifically, we compute
\begin{equation}
    \bar{R}_\alpha(\ell)
    =
    \left\langle R_\alpha^\alpha \right\rangle_e^{1/\alpha}
    \quad
    \text{and}
    \quad
    \bar{C}(q)
    =
    \left\langle C(q) \right\rangle_e
\end{equation}
where $\langle\cdot\rangle_e$ is now a time/ensemble-average over different realizations of the system and the four partials, obtained in snapshots written at $1$~ps intervals during the isothermal simulation run.
This protocol was applied to characterize the dislocation structure at different finite temperatures, from \SI{5}{\kelvin} to \SI{1000}{\kelvin}.

\begin{figure}[t]
 \includegraphics[clip,width=\columnwidth]{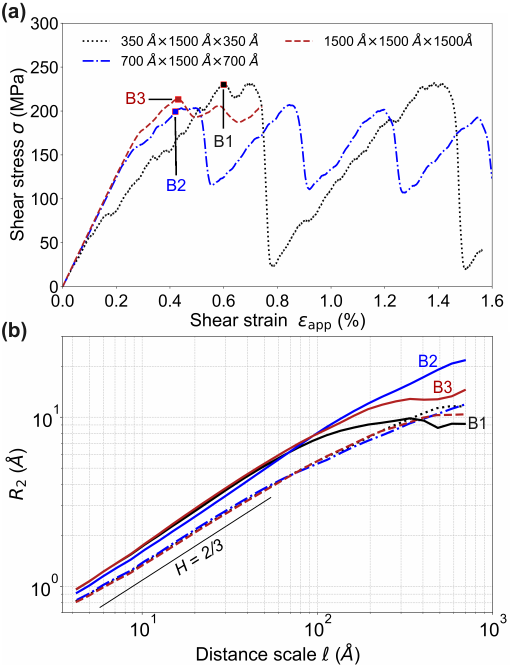}
\caption{\label{fig:stress_strain_size_effect}\textbf{Phenomenology of plastic flow and dislocation structure before and after depinning at $\textbf{5}$~K.}  \textbf{(a)} Shear stress as a function of applied shear strain for different sizes of the simulation cell. \textbf{(b)} Autocorrelation function $R_2(\ell)$ of the dislocation position.
Dashed lines show the dislocations at zero stress, while
solid lines show the structure before depinning (at points B1, B2, B3 in panel (a)).}
\end{figure}

\section{Results} \label{results}

We start with the phenomenology of plastic flow in our model setup of the FeNiCrCoCu alloys using well-equilibrated dislocations.
Figure~\ref{fig:stress_strain_size_effect}a shows the shear stress $\sigma$ as a function of the applied shear strain $\varepsilon_\text{app}$ in the HEA during shear of the simulation boxes at \SI{5}{\kelvin}.
At small strain, the system behaves elastically, $\sigma=G \varepsilon_\text{app}$ with shear modulus $G$, up to $1.7\%$ strain.
Deviation from linearity and serrations in the stress-strain response indicates the onset of plastic flow.
The plastic regime shows macroscopic depinning events in which stress drops at repeating intervals.
This depinning is due to the mutual interaction of the two dislocations with opposite Burger's vector in the dipole, which pass each other in these events.
Because the dislocations then move through the periodic boundaries and meet again, the stress-strain response appears periodic as the simulation box continues to shear.

Figure~\ref{fig:stress_strain_size_effect}a shows results obtained for different sizes of the simulation cell.
All simulations have dislocations of identical length, but the perpendicular directions are varied leading to different spacing of the two dislocations in the dipole.
As expected, the dislocation spacing has an effect on the stress-strain response, affecting in particular the macroscopic depinning events.
Only the simulation with a cubic box of $(\SI{1500}{\AA})^3$ does not appear to show dominant macroscopic depinning, i.e. the two dislocations appear far enough such that their interaction becomes negligible.

During deformation, we extract average position and structure of each dislocation in our simulations using the dislocation extraction algorithm (DXA)~\cite{Stukowski2010ExtractingData}.
The dislocations adopt wavy configurations (see Fig.~\ref{fig:sim_box}c), which change constantly during movement through the fluctuating chemical composition.
This leads to microscopic depinning events, visible by smaller serrations in the stress-strain curve.
We specifically record the dislocation structure at zero shear stress and before the largest (macroscopic) depinning events, where stresses are maximal, indicated by B(efore)1, B2, B3 in Fig.~\ref{fig:stress_strain_size_effect}a.

Figure~\ref{fig:stress_strain_size_effect}b shows the correlation function $R_2(\ell)$ at these points. 
The power-law correlations at zero stress (dashed lines in Fig.~\ref{fig:stress_strain_size_effect}b) and right before depinning (B1, B2, B3, solid lines in Fig.~\ref{fig:stress_strain_size_effect}b) show identical scaling at short distances, but the overall amplitude of $R_2$ is higher right before depinning.
This means the dislocations exhibit a configuration that appears wavier at high stress.
Note that the structure of the dislocation is not strongly dependent on the dipole spacing:
Although the distance between the dislocations in the dipole affects the shear stress-strain response, it does not impact the roughness of the dislocation.
All further results will, therefore, be shown on the smaller system with a periodic dimension of $350$~Å in the $[1 \bar{1} 0 ] $ and $[1 1 1]$ directions perpendicular to the dislocation lines.
\begin{figure}[t]
  \includegraphics[clip,width=\columnwidth]{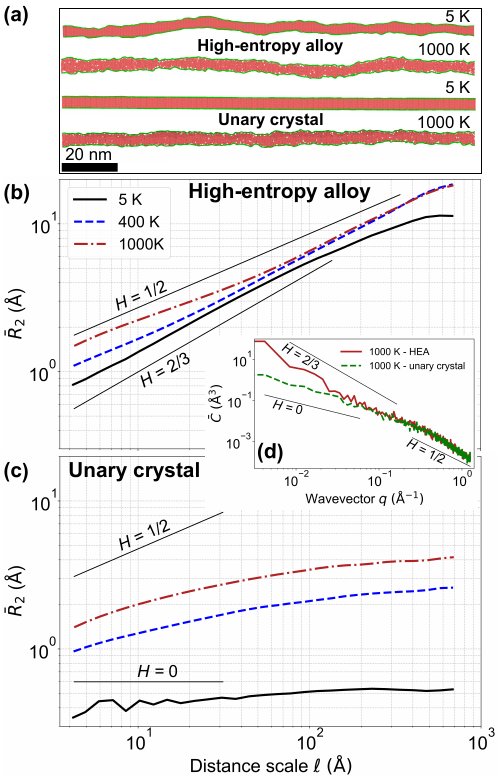}%
  \caption{\label{fig:sqrtACF_HEA_finite_temp}\textbf{Temperature dependence of 
 the dislocation structure at zero stress.}
 \textbf{(a)} Snapshots of dislocations in the HEA and an unary crystal in the AA-model at \SI{5}{\kelvin} and \SI{1000}{\kelvin}.
 The autocorrelation function $\bar R_2(\ell)$ of these dislocations are shown in panel \textbf{(b)} for the HEA and in panel \textbf{(c)} for the unary crystal in the AA-model. The figure shows results at \SI{5}{\kelvin}, \SI{400}{\kelvin} and \SI{1000}{\kelvin}. Each curve is obtained by averaging the autocorrelation function $R_2(\ell)$ over $10$ snapshots. Panel \textbf{(d)} shows the plot of the power spectra density $\bar C(q)$ of HEA and unary crystal at \SI{1000}{\kelvin}, obtained by averaging the PSD $C(q)$ over 10 snapshots. Thin solid lines show ideal power-law behavior $\bar R_2(\ell)\propto \ell^H$ or $\bar C(q)\propto q^{-1-2H}$ for the Hurst exponent $H$ indicated in the figure.}

\end{figure}

We now examine the dislocation structure in thermal equilibrium at zero shear stress in slowly quenched systems.
Figure~\ref{fig:sqrtACF_HEA_finite_temp}a shows the snapshots of two Shockley partial dislocations separated by stacking faults (red color areas) in the HEA and the average alloy at \SI{5}{\kelvin} and \SI{1000}{\kelvin}.
Even at low temperatures (\SI{5}{\kelvin}), the dislocations in the HEA exhibit wavy configurations, which become rougher with thermal activation.
This observation aligns with the measurement of the ensemble-averaged autocorrelation function $\bar R_2(\ell)$ shown in Fig.~\ref{fig:sqrtACF_HEA_finite_temp}b, which shows an increase in amplitude at low distances $\ell$ as temperature increases.

The autocorrelation function $\bar R_2(\ell)$ contains more information on the spatial structure of the dislocation.
A power-law scaling $\bar R_2(\ell)\propto \ell^H$ with $0<H<1$ indicates a self-affine morphology with Hurst~\cite{Hurst1951Long-TermReservoirs,Mandelbrot1982-wd} exponent $H$.
Figure~\ref{fig:sqrtACF_HEA_finite_temp}b shows trend lines obtain for ideal power-laws with $H=1/2$ (random walk) and $H=2/3$.
At low temperature (\SI{5}{\kelvin}), the dislocation line in the HEA is power-law correlated with a Hurst exponent $H \approx 2/3$.
When the temperature increases, we observe a transition to a Hurst exponent $H \approx 1/2$, indicative of a random walk. 
At larger distances $\ell$, the simulations at the higher temperatures ($400$~K and $1000$~K) appear to cross over to $H\approx 2/3$.

In the unary crystal, which has identical elastic properties but no disorder, dislocations are straight at low temperature. 
Similar to the HEA, thermal effects cause dislocations to fluctuate and become rougher, as illustrated in Fig.~\ref{fig:sqrtACF_HEA_finite_temp}a.
Visual comparison of the high-temperature dislocation in the HEA and the unary crystal already hints that their larger-scale structures differ.
In Fig.~\ref{fig:sqrtACF_HEA_finite_temp}c, we show $\bar R_2(\ell)$ of the unary crystal.
The autocorrelation is flat at low temperature (\SI{5}{\kelvin}), in accordance with the straight dislocation in Fig.~\ref{fig:sqrtACF_HEA_finite_temp}a.
At \SI{1000}{\kelvin}, the autocorrelation remains flat at a large distance $\ell$, but appears to cross over to an exponent $H \approx 1/2$ at a small distance. 

We also compare the ensemble-averaged power-spectral density (PSD) $\bar C(q)$ of the HEA and the unary crystal at \SI{1000}{\kelvin}, shown in Fig.~\ref{fig:sqrtACF_HEA_finite_temp}d.
For a self-affine morphology, the PSD also is a power law $\bar C(q)\propto q^{-1-2H}$.
Consistency of the exponent $H$ between real-space measurements $\bar R_2(\ell)$ and reciprocal-space measurements $\bar C(q)$ is a strong indicator of self-affinity.
At small $q$, $\bar C(q)$ of the HEA shows power law behavior with a Hurst exponent $H \approx 2/3$, while the unary crystal shows uncorrelated white noise. At large $q$, $\bar C(q)$ of the HEA and the unary crystal show power-law scaling with $H \approx 1/2$.
This observation is consistent with the measurements obtained from the autocorrelation function $\bar R_2(\ell)$. 
\begin{figure}[t]
  \includegraphics[clip,width=\columnwidth]{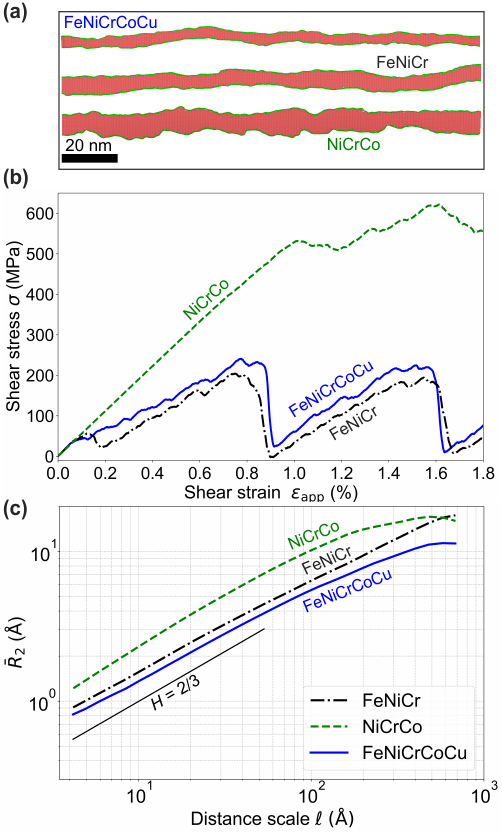}
  \caption{\label{fig:compare alloys}\textbf{MD simulation results of plastic flow and dislocation structure at varying alloy composition.} \textbf{(a)} Snapshot of dislocations in FeNiCrCoCu, FeNiCr, NiCrCo equicomposition HEAs at 5 K and zero stress. \textbf{(b}) Shear stress as a function of applied shear strain for NiCrCo (green), FeNiCr (black) and FeNiCrCoCu (blue). \textbf{(c)} Autocorrelation function of dislocations  $\bar{R}_2(\ell)$ in NiCrCo (green), FeNiCr (black) and FeNiCrCoCu (blue) equicomposition HEAs at 5K and zero stress.}
  
\end{figure}

We performed similar MD simulations for FeNiCr and NiCrCo equicomposition alloys. Differences in well-equilibrated dislocation geometries are apparent in Fig.~\ref{fig:compare alloys}a. In NiCrCo, the two Shockley partial dislocations exhibit rougher profiles and show a larger stacking fault distance than FeNiCrCoCu and FeNiCr.  NiCrCo exhibits significantly higher shear stress than the other alloys (Fig.~\ref{fig:compare alloys}b) and is the only alloy that does not show a pronounced effect of the dislocation dipole setup on the stress-strain curve.
The dislocations in all three alloys (see $R_2(\ell)$ in Fig.~\ref{fig:compare alloys}c) appear to have a fractal structure with a Hurst exponent $H\approx 2/3$, with slight variations between the alloys.
The structure appears to become uncorrelated at long distances in NiCrCo and FeNiCrCoCu; such decorrelation is not observable for FeNiCr in the simulation cell sizes studied by us.

\begin{figure}[t]
  \includegraphics[clip,width=\columnwidth]{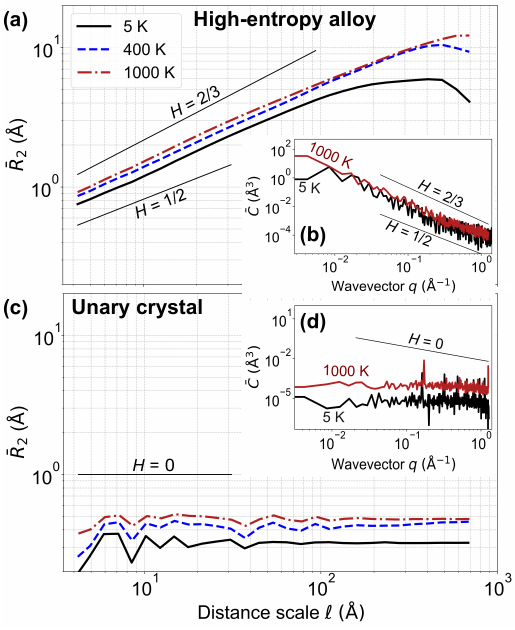}%
\caption{\label{fig:sqrtACF_IH}\textbf{Inherent structure of dislocations at zero stress.} The inherent structure of the system is obtained by an instantaneous quench of a system with two straight dislocations to the closest local minimum. The system was equilibrated at the indicated temperature before the quench. The figure shows the autocorrelation function $\bar R_2(\ell)$ and power spectral density $\bar C(q)$. Panel \textbf{(a)} and panel \textbf{(b)} show results of $\bar R_2(\ell)$ and $\bar C(q)$ for the HEA at parent temperatures 5 K, 400 K, 800 K and 1000 K;
panel \textbf{(c)} and panel \textbf{(d)} show results of $\bar R_2(\ell)$ and $\bar C(q)$ for an unary crystal in the AA model at a parent temperature of 5K and 1000 K.}

\end{figure}

We now address the effect of thermal fluctuations by studying the inherent structure of the finite temperature systems above.
The inherent structure removes the influence of thermal fluctuations but keeps the influence of quenched disorder.
Figure~\ref{fig:sqrtACF_IH}a shows the $\bar R_2$ of the inherent structure in FeNiCrCoCu at different parent temperatures.
The structure obtained at the higher parent temperatures of $400$~K and $1000$~K appears to be similar, while the lower parent temperature of $5$~K has an overall smaller amplitude.
Additionally, the higher parent temperature appear to scale with $H\approx 2/3$ while the lower parent temperature has a smaller exponent of around $H\approx 1/2$.
The PSD (Fig.~\ref{fig:sqrtACF_IH}b) shows power-law scaling with exponents consistent with this observation.
Conversely, the unary alloy shows no structure.
Both the autocorrelation function (Fig.~\ref{fig:sqrtACF_IH}c) and the PSD (Fig.~\ref{fig:sqrtACF_IH}d) are flat, indicating white noise.
The amplitude of the autocorrelation function is below $1$~\AA, meaning the dislocation is essentially straight.

\begin{figure}[t]
  \includegraphics[clip,width=\columnwidth]{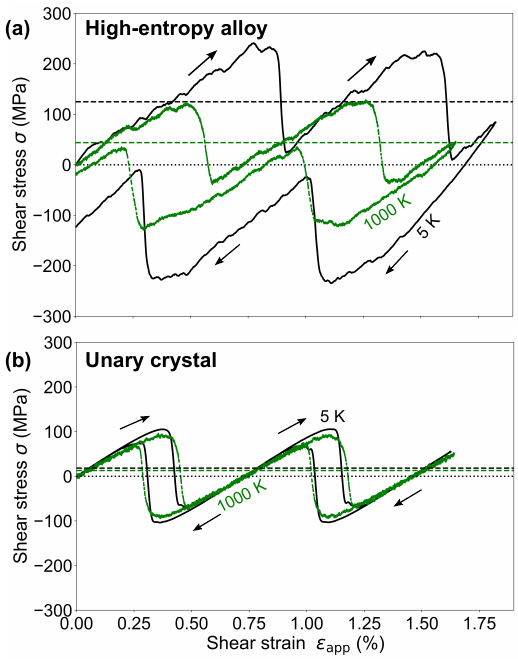}%
\caption{\label{fig:shear_finite_temp}\textbf{Hysteresis and energy dissipation during dislocation motion.} The figure shows the shear stress vs the applied shear strain for \textbf{(a)} the HEA and \textbf{(b)} the unary crystal in the AA-model at \SI{5}{\kelvin} and \SI{1000}{\kelvin}. The simulation box was sheared up to roughly $1.7\,\%$ before reversing the shear direction. The dashed lines show the average shear stresses within one macro-depinning cycle. The area between the stress-strain curves of the two shearing directions is the energy dissipated in the deformation loop. The HEA shows higher energy dissipation than the AA because the dislocation line is pinned by the compositional disorder. Pinning becomes weaker at high temperatures, indicated by a contraction of the hysteresis loop. For the unary crystal, temperature does not affect the overall dissipation.}
\end{figure}

Finally, we look into dislocation glide at finite temperatures, starting with a phenomenological point of view.
We shear the simulation box at a strain rate of $10^{7}\,\text{s}^{-1}$ up to $\sim 1.7\%$ strain before reversing the shear direction. 
The resulting stress-strain curves at \SI{5}{\kelvin} and \SI{1000}{\kelvin} in the HEA are shown in Fig.~\ref{fig:shear_finite_temp}a.
The initial quasi-elastic response extends to higher stress (and larger strain) as the temperature decreases in the HEA.
This is indicative of stronger pinning of the dislocations at lower temperature. 
Reversing the shear direction allows us to study hysteresis in the stress-strain response.
Hysteresis means dissipated energy, which is proportional to the area within the hysteresis loop in a stress-strain curve.
The hysteresis loop is wider at low temperature, indicating stronger dissipation.
The behavior of the unary crystal differs qualitatively from the HEA.
Figure~\ref{fig:shear_finite_temp}b shows a stress-strain response that is identical for low and high temperature.
The hysteresis is significantly smaller than what we observe for the HEA, with a small hysteretic loop remaining around the depinning of the dipole array.
We note that hysteresis means that the average stress for moving the dislocation is nonzero, e.g. on the order of $100$~MPa for the dislocation in the HEA at $5$~K shown in Fig.~\ref{fig:shear_finite_temp}a.
Conversely, the average stress to move the dislocation in the unary crystal is close to zero.
As temperature increases, the dislocation finds it easier to depin, leading to a contraction of the hysteresis loop.
For the unary crystal, the overall dissipation remains unaffected by temperature.

\section{Discussion} \label{sec:discussion}

We begin by examining the influence of the periodic dipole used in this study.
Periodic boundary conditions eliminate surfaces and preserve the crystal lattice's translational symmetry, but dislocations interact through their elastic deformation fields in the periodic dislocation array.
Interaction forces can be straightforwardly quantified using well-known expressions from the linear theory of elasticity \cite{Hirth1983TheoryEd.,Hull2011IntroductionDislocations}.
For example, Cai and co-workers \cite{Cai2003PeriodicModelling, Kuykendall2013ConditionalDynamics} 
have simply summed contributions of repeating unit cells, where care must be taken of the summation order because -- like in the electrostatic case -- the sum is conditionally convergent. 
Using the mean dislocation positions extracted using DXA from the molecular dynamics calculations, we computed the force $F_\text{disl}$ on each dislocation by summing contributions from twelve nearest neighbors.
For each pair we use the canonical expression for the force between two dislocations in an isotropic elastic medium~\cite{Hirth1983TheoryEd., Hull2011IntroductionDislocations},
\begin{equation}
    F_\text{pair}(d,p) = -\frac{Gb^2 d (d^2-p^2)}{2\pi (1-\nu)(d^2+p^2)^2},
    \label{eq:peach-koehler}
\end{equation}
where $d$ is the distance between the dislocation in the glide direction and $p$ perpendicular to it (see Fig.~\ref{fig:sim_box}).
Note that Eq.~\eqref{eq:peach-koehler} is the expression for isotropic elasticity, while our system is anisotropic.
For the shear modulus $G$ we therefore use the value obtained in the direction of simple shear that is imposed in our simulations.
For unary crystals, the dislocation is pinned by the Peierls-Nabarro barrier, but in HEAs there is additional pinning because of fluctuations in chemical composition
This is typically expressed by a pinning field $U_\text{p}$ in the quenched EW-Model, Eq.~\eqref{eq:qew}.
Utt and co-workers~\cite{Utt2022TheAlloys} extracted this pinning field directly from atomistic calculations.
The force is balanced by the applied shear stress $\sigma$, such that the Peach-Köhler force $F_\text{disl} = \sigma b$ with Burger's vector $b$, allowing us to compute $\sigma$ from the dislocation position.

\begin{figure}
 \includegraphics[clip,width=\columnwidth]{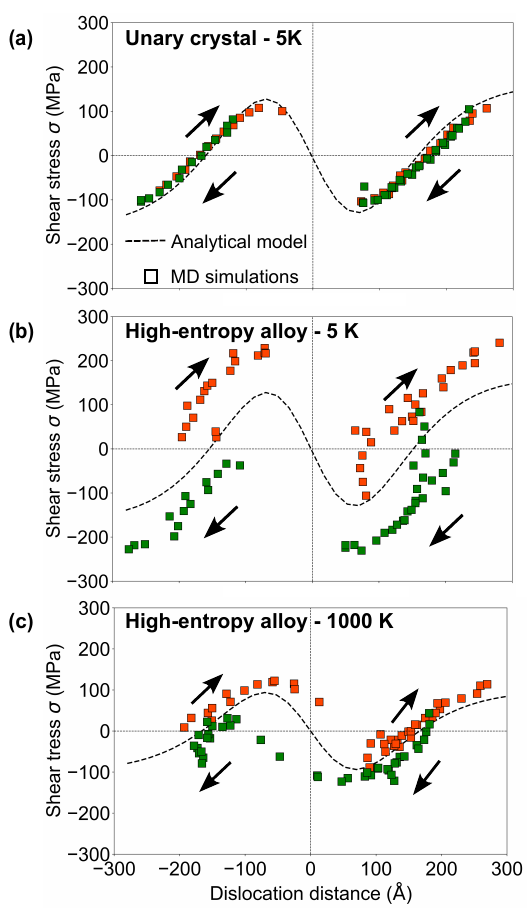}
\caption{ \label{fig:dis_analytical_model}\textbf{Continuum model of interaction stresses in an edge dislocation dipole.} The analytical model illustrates interaction stresses for passing dislocation (dashed lines) and MD simulation results (orange and black points). Panel \textbf{(a)} shows the result for the unary crystal in the AA-model at \SI{5}{\kelvin}, panel \textbf{(b)} for the HEA at \SI{5}{\kelvin} and panel \textbf{(c)} for the HEA at \SI{1000}{\kelvin}. The analytical model fails to capture the interaction of the dislocations in the low-temperature HEA.}
\end{figure}

In Fig.~\ref{fig:dis_analytical_model}, we show $\sigma$ from this simple elastic interaction model as a function of the distance of the two dislocations in the unit cell.
We also show the shear stress extracted directly from our MD simulations using the canonical virial expression~\cite{Allen1989-nt}.
At low temperature ($5$~K, Fig.~\ref{fig:dis_analytical_model}a), dislocations in the unary crystal are mostly straight, like in a pure material and the model exactly reproduces the shear stress from our calculations.
Conversely, in the HEA, there is strong pinning by chemical disorder and the stress in the full molecular calculation is significantly higher than in the model (Fig.~\ref{fig:dis_analytical_model}b), with a difference on the order of $100$~MPa.
This difference completely disappears at higher temperature (Fig.~\ref{fig:dis_analytical_model}c).
This is consistent with the observation of hysteresis in the stress-stress curves shown in Fig.~\ref{fig:shear_finite_temp}, where the low-temperature HEA shows a yield stress around $100$~MPa higher than the high-temperature HEA and the unary alloy.
We conclude that the high-temperature calculations are not pinned, meaning that thermal fluctuation essentially smooth out most of the variation in pinning field that the dislocation sees.

We now connect these thermodynamic properties back to the structure of the line.
Without quenched disorder, thermal fluctuations yield dislocation lines with a Hurst exponent of $1/2$ at long time scales~\cite{Geslin2018ThermalElasticity,Zhai2019PropertiesDisorder}.
To understand this behavior, it is instructive to look at the time evolution of the dislocation line, which is often modeled as Ginzburg-Landau type (gradient flow) dynamics of Eq.~\eqref{eq:qew} (e.g. Ref.~\cite{Zaiser2002DislocationSolution}),
\begin{equation}
    B \frac{\partial h}{\partial t}
    =
    -\frac{\delta A}{\delta h(x)}
    =
    \Gamma \frac{\partial^2 h}{\partial x^2} + F_\text{p}(x, h(x,t)) + f
    \label{eq:qew-time-dependence}
\end{equation}
where $\delta/\delta h(x)$ is a functional derivative.
Here, $B$ is a dynamic (damping) coefficient and $F_\text{p}=-\partial U_\text{p}/\partial h$ the pinning force.
In Fourier space, the Green's function of Eq.~\eqref{eq:qew-time-dependence} scales like $\propto q^{-2}$ with wavevector, which is exactly the scaling observed in the PSD shown in the inset to Fig.~\ref{fig:sqrtACF_HEA_finite_temp} and corresponds to a Hurst exponent $H=1/2$~\cite{Meakin1998-mw}.
This scaling has been observed in a Langevin formulation of Eq.~\eqref{eq:qew-time-dependence}, sometimes called the annealed EW-model~\cite{Zhai2019PropertiesDisorder}.

We recover this scaling in our molecular dynamics calculations of the HEA at short distances (see $1000$~K results in Fig.~\ref{fig:sqrtACF_HEA_finite_temp}b).
Rising temperature increases the amplitude of the correlation function $\bar{R}_2$, and as a consequence, the fluctuation-induced structure of the dislocation extends to larger distances.
This is in accordance with the observation of the disappearance of pinning in the high-temperature HEA calculations.
However, the arguments above and numerical calculations~\cite{Zhai2019PropertiesDisorder} indicate that the power-law region (with Hurst exponent $1/2$) should extend to all distances at large times.
In our calculations of the unary crystal, $\bar{R}_2$ does not change when equilibrating the calculation for $2$~ns, indicating that the dislocation line in these calculations will not converge to the structure of a random walk.
This appears consistent with molecular dynamics results from Ref.~\cite{Geslin2018ThermalElasticity}, who argued that there should be a logarithmic correction to the elastic kernel which is absent in the EW model.
However, the high-temperature HEA calculations (which are not pinned and show only negligible hysteresis) appear to follow a power-law (with an exponent of roughly $1/2$) up to the system size.
Note that if we quench the system instantaneously to the next local minimum to compute the inherent structure, the dislocation in the unary alloy becomes straight while the HEA shows power-law correlation with $H\approx 2/3$ at high parent temperature (see Fig.~\ref{fig:sqrtACF_IH}).
For a dislocation that is created perfectly straight and then equilibrated at a low temperature, we find a Hurst exponent of $H\approx 1/2$ and not $2/3$ (see $5$~K results in Fig.~\ref{fig:sqrtACF_IH}a).
However, when the dislocation is created by slowly quenching from a high temperature, we see $H\approx 2/3$ at low-temperature (see $5$~K results in Fig.~\ref{fig:sqrtACF_HEA_finite_temp}b).

At low temperatures, chemical heterogeneity traps the dislocations, with individual atoms acting as obstacles to dislocation glide.
%
%
Dislocations bow out at a characteristic scale $\lambda_\text{p}$ that represents a balance between the line tension $\Gamma$, which tries to straighten the line, and the applied shear stress, which allows the segment to bow out.
A classic argument~\cite{Larkin1970-gy,Larkin1979-rv,Robbins1987-kh,Zhai2019PropertiesDisorder} goes as follows:
A bulge with displacement $\xi$ over length $\lambda_\text{p}$ increases the length of line by $(\xi/\lambda_\text{p})^2 \lambda_\text{p}$ and hence costs energy $\Delta E_\text{bulge} \sim \Gamma (\xi/\lambda_\text{p})^2 \lambda_\text{p}$.
This will yield a restoring force
\begin{equation}
    F_\text{bulge} \sim \Delta E_\text{bulge}/\xi \sim \Gamma \xi/\lambda_\text{p}.
\end{equation}
where $\xi$ is the characteristic length over which the force varies, i.e. the correlation length of the underlying random field.
For a random pinning field with characteristic energy (per unit length) $U_0$, the characteristic force is $U_0/\xi$.
This means the characteristic force over the bulge with area $\xi L$ is
\begin{equation}
    F_\text{random} \sim U_0/\xi \sqrt{\xi L} = U_0 \sqrt{L/\xi}.
\end{equation}
From $F_\text{bulge}=F_\text{random}$ we get
\begin{equation}
    \lambda_\text{p}/\xi = (\Gamma/U_0)^{2/3}.
\end{equation}
Dislocations navigate through the crystal by propagating stiff segments length $\lambda_\text{p}$ of the dislocation line. 

\begin{figure}
\centering
 \includegraphics[width=\columnwidth]{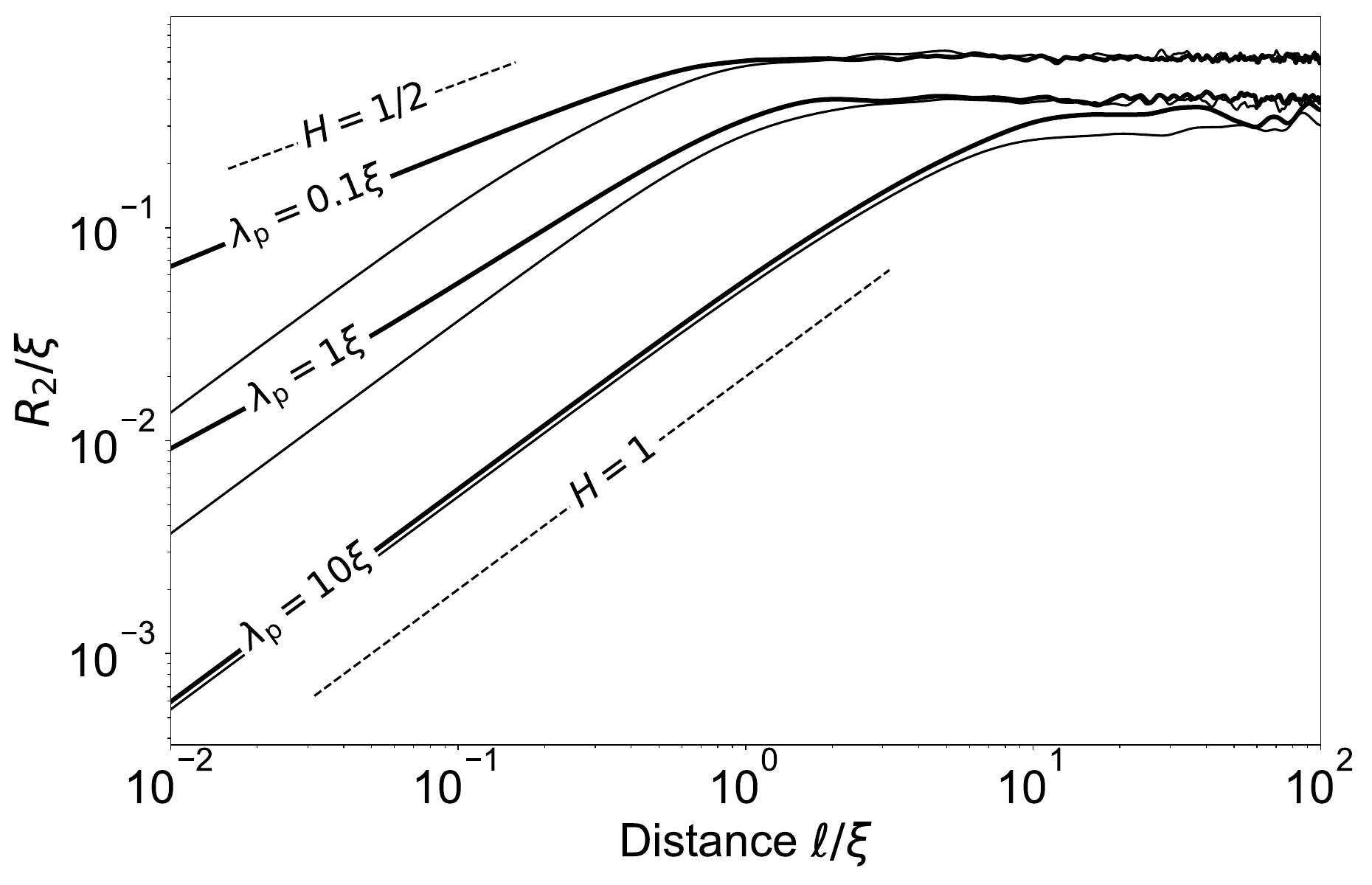}%
\caption{\label{fig:qew}\textbf{Numerical solution of the line-tension model.} Simulations of the dislocation structure in the quenched arclength EW model (thick lines) show different scaling exponents for stiff lines ($\lambda_\text{p}>\xi$) and floppy lines ($\lambda_\text{p}<\xi$). For comparison we show results obtained with the standard, linearized EW model (thin lines). The linearized model shows $H=1$ in all cases. The simulations started with a straight dislocation that was relaxed to the nearest local minimum in free energy.} 
\end{figure}

Only our low-temperature calculation ($5$~K) show clear pinning and hysteresis.
In these simulations, we observe a Hurst exponent of $H$ between $1/2$ and $2/3$, where the latter result is obtained only if the dislocation is well-equilibrated at a high temperature.
We note that this result appears to be independent of the alloy:  Figure~\ref{fig:compare alloys}c shows $\bar{R}_2$ for well-equilibrated dislocations in three different alloys that all appear to yield the same scaling exponent.
We note that this is consistent with prior MD simulations~\cite{Peterffy2020LengthSolutions}, but inconsistent with numerical solutions of the quenched EW-model, Eq.~\eqref{eq:qew}, which show a Hurst exponent of $1$~\cite{Bako2008DislocationSteel,Zhai2019PropertiesDisorder}.
We carried out our own calculations (see numerical details in Appendix~\ref{app:num-qew}) on a random field with correlation length $\xi$ and can confirm this scaling for all $\lambda_\text{p}$.
However, the free-energy described by Eq.~\eqref{eq:qew} is only proportional to the length of the dislocation for small $\partial h/\partial x$.
A generalized model uses the arc-length expression (see e.g. Ref.~\cite{Geslin2020InvestigationParameters}), yielding
\begin{equation}
    A_\text{arc} = \int_0^L \dif x \, \left\{\Gamma \left[\sqrt{1+\left(\partial_x h\right)^2}-1\right]+U_\text{p}(x, h) - f h\right\},
    \label{eq:qew-arclength}
\end{equation}
which is identical to Eq.~\eqref{eq:qew} for small $\partial h/\partial x$.
Figure~\ref{fig:qew} shows that our numerical simulations show $H=1$ only for stiff lines ($\lambda_\text{p}>\xi$) where the EW model is a good approximation, but they yield $H=1/2$ for floppy lines ($\lambda_\text{p}<\xi$) in the individual pinning limit.
The power-law region crosses over to constant correlation at a length-scale of $\lambda_\text{p}$ for stiff lines and $\xi$ for floppy lines.
We note that our EW calculations start from a straight line which we quench to the next local minimum.
This means these results need to be compared with our MD calculations of the inherent structure, Fig.~\ref{fig:sqrtACF_IH}, which also show $H=1/2$.

Our MD simulation also show a crossover to constant $\bar{R}_2$, whose transition length is on the order of $5-50$~nm but appears to depend on alloys (see Fig.~\ref{fig:compare alloys}).
Prior work has reported that this length growth close to depinning~\cite{Zhai2019PropertiesDisorder} and we confirm this observation in our EW calculations.
Since the exponent is not unity, our arclength EW calculation indicate that this length scale must be the intrinsic correlation length of the underlying disorder field.
The chemical disorder in our HEAs is uncorrelated; a naive expectation would therefore be that the instrinsic correlation length is on the order of the lattice constant ($\sim 0.4$~nm) of the crystal. 
This seems to indicate that there may be intrinsic correlations in the effective disorder that the dislocation sees, maybe due to microdistortions, in the HEA.
Indeed, Nöhring and Curtin speculate~\cite{Nohring2019-qq}, that such microdistortions are responsible for the breakdown of the solute-strengthening theory of HEAs~\cite{Varvenne2017SoluteAlloys}.
We note that another candidate for an intrinsic length-scale is the splitting distance of the two Shockley partials, but the alloy with the largest splitting distance (NiCrCo, Fig.~\ref{fig:compare alloys}a) actually shows the shortest cross-over length (Fig.~\ref{fig:compare alloys}b).

\section{Summary \& Conclusions}

In summary, we conducted atomistic simulations to investigate the dislocation structures under the combined effects of quenched and thermal disorder in HEAs.
In line with previous findings, dislocations in HEAs exhibit correlated profiles with a Hurst exponent of $H \approx 2/3$ at low temperature, but only if the dislocation is well-equilibrated.
For non-equilibrated, straight dislocations that are quenched to the next local minimum, we find $H=1/2$.
Calculations of a simple nonlinear line-tension model confirm $H\approx 1/2$, but only a regime where the line is ``floppy''.
This indicates that the dislocations in the three alloys studied here are in a regime where the Larkin length of the line is smaller than the intrinsic correlation length of the disorder field.

High-temperature results differ significantly between unary crystals and HEA: Both show $H \sim 1/2$ at low temperature but only in the HEA this scaling extends across all scales. 
The transition from low- to high-temperature corresponds to the loss of dislocation pinning, evidenced by the disappearance of hysteresis in dislocation motion. 
Our results indicate that dislocations in HEAs may be in an individual pinning limit, where segments of the dislocation are independently pinned by local distortions of the crystal lattice induced by chemical heterogeneity.
The results important for the development of an analytic theory of strengthening in HEAs.

\begin{acknowledgments}
We thank Patrick Dondl, Péter Ispánovity and Antoine Sanner for useful discussions.
Funding was provided by the Deutsche Forschungsgemeinschaft (DFG) within SPP2256 (grant 441523275) and the European Research Council (StG 757343).
Calculations were run on NEMO (University of Freiburg, DFG grant INST 39/963-1 FUGG) and HoreKa (Karlsruhe Institute of Technology, project ``HEADislocation'').
We used \textsc{lammps}~\cite{Thompson2022LAMMPSScales} for all molecular dynamics calculations reported here.
We preprocessed, postprocessed and visualized atomic simulation data with \textsc{ovito}~\cite{Stukowski2010ExtractingData} and \textsc{ASE}~\cite{Larsen2017TheAtoms}.
\end{acknowledgments}

\appendix

\section{Line-tension model}
\label{app:num-qew}

We discretize the energies Eq.~\eqref{eq:qew-arclength} in a periodic domain using canonical linear finite elements with equal node spacing $\Delta x$.
For the integration of nonlinear terms we use Gaussian quadrature with a single quadrature point.
This yields the discrete energy
\begin{equation}
\begin{split}
    A_\text{arc}({h_i}) =& \Delta x \sum_i \left\{
        \Gamma \left[\sqrt{1+\left(\frac{h_{i+1}-h_i}{\Delta x}\right)^2}-1\right]
        \right.
        \\
        &\qquad\qquad\left.+
        U_\text{p}\left(x_{i+1/2}, h_{i+1/2}\right)
        -
        f h_i,
    \right\}
\end{split}
\end{equation}
where $h_i=h(x_i)$ the discrete heights.
A respective linearized equation holds for Eq.~\eqref{eq:qew}. 
Here, a half-index indicates the mean value, i.e. $h_{i+1/2}=(h_i+h_{i+1})/2$.
The random pinning field $U_\text{p}$ has rms amplitude $U_0$, and we created it by Fourier-filtering white noise such that the power of the pinning field for all wavevectors larger than $2\pi/\xi$ is zero.
For static calculations, we find the nearest local minimum of $A({h_i})$ using the L-BFGS algorithm starting from a straight line.


%

\end{document}